\documentclass[twocolumn,superscriptaddress,floatfix, amsfonts, aps]{revtex4-1}
\usepackage{amsmath}
\usepackage{bm}
\usepackage{graphicx}
\usepackage{subfigure}
\usepackage{float}
\usepackage{mathrsfs}
\usepackage{footmisc}
\usepackage{multirow}
\usepackage{graphicx}
\usepackage{soul,xcolor}
\usepackage{xcolor, ulem, color, framed}
\usepackage{amssymb}
\usepackage{setspace}
\usepackage[colorlinks, linkcolor=red,anchorcolor=red,citecolor=blue]{hyperref}

\newcommand{\tred}[1]{{\color{black}\textbf{}\,{#1}}}
\graphicspath{{./figs-pp/}}

\begin{document}

\title{Probing QGP droplets with charmonium in high-multiplicity proton-proton collisions}

\author{Yunfan Bai}
\affiliation{Department of Physics, Imperial College London, London SW7 2AZ, United Kingdom}

\author{Baoyi Chen}\email{baoyi.chen@tju.edu.cn}
\affiliation{Department of Physics, Tianjin University, Tianjin 300354, China}

\date{\today}

\begin{abstract}
We study the hot medium effects in high-multiplicity proton-proton (pp) collisions at $\sqrt{s_{NN}}=13$ TeV via the charmonium probes. The hot medium is described with the hydrodynamic model, while charmonium evolutions in the medium are studied with a time-dependent Schr\"odinger equation. The hot medium dissociation on charmonium is considered with the temperature-dependent complex potential parametrized with the results from lattice QCD calculations. The ratio $\sigma_{\psi(2S)}/\sigma_{J/\psi}$ of $J/\psi$ and $\psi(2S)$ production cross sections are calculated and compared with the LHCb experimental data in pp collisions. Our calculations explain the charmonium relative suppression in different transverse momentum and multiplicity bins. The suppression of this ratio is mainly affected by the effects of the deconfined medium. It is less affected by the initial effects before the generation of the heavy quark pair. We suggest this to be a clear signal of the small QGP droplets generated in high multiplicity pp collisions.

\end{abstract}
\maketitle

\section{Introduction}
In ultra-relativistic heavy-ion collisions, deconfined matter is believed to form in the early stages~\cite{Bazavov:2011nk}. This deconfined matter primarily consists of partons such as quarks and gluons, called Quark-Gluon Plasma (QGP), akin to the early Universe after the Big Bang. 
Approximately forty years ago, Matsui and Satz proposed that charmonium's anomalous suppression could indicate the existence of the deconfined matter in ultra-relativistic heavy-ion collisions~\cite{Matsui:1986dk}. The fundamental concept behind this proposal is that the color screening effect from thermal partons within QGP reduces the binding energy of charmonium, leading to the melt of charmonium states~\cite{Mocsy:2007jz,Chen:2013wmr,Du:2017qkv,Rothkopf:2019ipj,Satz:2005hx,Liu:2012zw}. The temperature at which charmonium binding energies drop to zero is called the dissociation temperature~\cite{Liu:2013kkg,Liu:2020cqa}. Below this temperature, parton inelastic scatterings can also dissociate charmonium states~\cite{Yan:2006ve,Aronson:2017ymv,Blaizot:2018oev,Yao:2020xzw,Wu:2024gil}. Consequently, heavy quarkonium experiences additional suppression in nucleus-nucleus collisions compared to proton-proton collisions~\cite{ALICE:2015kgk,CMS:2022wfi,ALICE:2019lga,ALICE:2018wzm,CMS:2017ycw}, where the presence of QGP was believed to be negligible.

Recent studies suggest a small QGP droplet may also form in small collision systems like p-Pb~\cite{Chen:2023toz,Chen:2023toz,Chen:2016dke,Du:2018wsj,Strickland:2024oat} and even in pp collisions~\cite{Weller:2017tsr,LHCb:2023xie,ALICE:2019zfl,Zhao:2020wcd}. Experiments have observed significant suppression of charmonium and bottomonium in p-Pb collisions $\sqrt{s_{NN}}=5.02$ TeV, phenomena well explained by QGP effects in theoretical models~\cite{Strickland:2024oat,Du:2018wsj,Chen:2016dke,Wen:2022utn}.
In recent high multiplicity proton-proton collisions at $\sqrt{s_{NN}}=13$ TeV, the ratio of charmonium production cross-sections $\sigma^{pp}_{\psi(2S)}/\sigma^{pp}_{J/\psi}$ is suppressed~\cite{LHCb:2023xie}. 
\tred{
As the initial cold nuclear matter effects occur prior to the production of heavy quark pairs ($gg \rightarrow c\bar{c}$) and the subsequent formation of charmonium states, these effects induce similar modifications in the production cross sections of charmonia due to the comparable masses of $J/\psi$ and $\psi(2S)$. Consequently, the ratio $\sigma^{pp}_{\psi(2S)}/\sigma^{pp}_{J/\psi}$ is primarily influenced by final state interactions with the hot nuclear matter that occur after the formation of $J/\psi$ and $\psi(2S)$. This makes the ratio of charmonium production cross sections a clean probe for deconfined matter in both small and large collision systems.
} 

\tred{In heavy-ion collisions, Schr\"odinger equation-based models have been developed to describe the dynamical evolution of heavy quarkonium~\cite{Strickland:2011mw,Islam:2020bnp,Wen:2022yjx,Katz:2015qja}. In these models, the Hamiltonian includes only the color-singlet sectors. The dissociation process reduces its survival probability and affects the normalization of the quarkonium wave packet in the medium. Consequently, inelastic scattering is parameterized with an imaginary potential in the Hamiltonian~\cite{Strickland:2011mw}. Additionally, the color screening effect is taken into account by incorporating a reduced real part of the potential. This potential model has successfully explained the suppression of heavy quarkonium in heavy-ion collisions~\cite{Islam:2020bnp,Wen:2022yjx,Guo:2015nsa}.

In small collision systems, we still utilize the Schr\"odinger model to investigate the suppression of heavy quarkonium in the hot medium~\cite{Song:2023zma}. First, the small hot medium may not achieve local equilibrium, and the fluctuations in energy density are significant in p-Pb and pp collisions; nonetheless, hydrodynamic models can still provide a good description of the final spectrum of light hadrons~\cite{Zhao:2020pty,Zhao:2017rgg,YuanyuanWang:2023meu}. Therefore, we adopt the temperature profiles given by the hydrodynamic model for the evolution of quarkonium. Second, the finite volume and the violent expansion of small collision systems may affect the properties of the small QGP and the decay of quarkonium~\cite{Liu:2020cqa,Guo:2018qxg,Wang:2024xeo,Pal:2023aai}. We continue to use the heavy quark potential extracted from lattice QCD results~\cite{Bala:2021fkm}, temporarily neglecting finite volume effects in pp collisions. Furthermore, the finite volume effect on the heavy quark potential can be partially reduced when examining the ratio of charmonium production cross sections as in this work.
}

Quantifying the contribution of QGP in pp collisions is critical for understanding quantum chromodynamics (QCD) interactions in finite-size systems and establishing baselines in physical observables for nucleus-nucleus collisions.
In this study, we utilize the Schr\"odinger equation model supplemented with a complex heavy quark potential to investigate the evolution of charmonium in high multiplicity pp collisions, where a small QGP is assumed to form and is characterized using a hydrodynamic model. Section II introduces the Schr\"odinger equation along with the in-medium heavy quark potentials. The QGP evolutions are also presented.
In Section III, we calculate the survival probabilities of $J/\psi$ and $\psi(2S)$ in the QGP droplets by employing various hydrodynamic backgrounds. The ratio of charmonium production cross-sections is also calculated and compared with the experimental data in different transverse momentum bins. In Section IV, a summary is given.

\section{Schr\"odinger equation and hydrodynamics}

\tred{
In heavy-ion collisions, heavy quarkonium experiences dissociation. Meanwhile, abundant heavy quarks are produced in initial parton hard scatterings and diffuse throughout the medium~\cite{Zhao:2017yan}. These quarks can combine to generate new quarkonium near the hadronization. Since regeneration production is proportional to the square of the heavy quark number and their dynamic evolutions~\cite{Andronic:2003zv,Greco:2003vf,Yan:2006ve,Yao:2018sgn,Du:2015wha}, this contribution becomes important in Pb-Pb collisions, but negligible in small collision systems such as p-Pb and pp collisions. 
Considering that anisotropic flows of D mesons have been observed in experiments~\cite{ALICE:2019fhe,ALICE:2022ruh}, which are partially induced by the expanding hot medium, we will study the evolution of primordial charmonium in pp collisions based on the assumption of a hot medium~\cite{Song:2023zma}. Different initializations of the medium energy density will be employed to explain charmonium suppression. 
}

In the potential model, the final state interactions of quarkonium are mainly referred to as the parton inelastic scatterings, parametrized as the imaginary potential in the Hamiltonian of heavy quarkonium.  
The relativistic corrections are neglected in the inner evolution of charmonium due to the large mass of charm quarks. With an isotropic in-medium heavy quark potential when the viscosity of the medium is taken as zero, the radial part of the Schr\"odinger equation is separated~\cite{Wen:2022utn,Wen:2022yjx} (in natural units $\hbar=c=1$), 
\begin{eqnarray}
\label{fun-rsch}
i {\partial \psi(r, t)\over \partial t}=[-{1\over 2m_\mu}{\partial ^2\over \partial r^2} +V(r,T) + {L(L+1)\over 2 m_\mu r^2}] \psi(r,t)
\end{eqnarray}
where $L$ denotes the angular quantum number,  $m_\mu = m_c/2$ represents the reduced mass in the center of mass frame of the heavy quark pair, with the charm quark mass $m_c = 1.5$ GeV. The function $\psi(r,t) = r R(r,t)$ is expressed as the product of the radius and the radial wave function, expandable in terms of $\psi(r,t) = r \sum_{nl} c_{nl}(t) \Phi_{nl}(r)$. $\Phi_{nl}(r)$ are the radial wave functions of charmonium eigenstates corresponding to the radial and angular quantum numbers $n$ and $l$. The coefficients $c_{nl}$ evolve due to the influence of the in-medium heavy quark potential $V(r,T)$.
To account for hot medium effects, a complex potential $V(r,T) = V_R(r,T) +V_I(r,T)$ is employed. The governing equation (Eq.(\ref{fun-rsch})) is numerically solvable using methods such as the Crank-Nicolson method~\cite{Kang:2006jd} or the Tridiagonal Matrix Algorithm~\cite{ref-trial}. Our simulations adopt spatial and temporal discretizations with steps of \( \Delta r = 0.03 \) fm and \( \Delta t = 0.001 \) fm/c, respectively. The choice of a computational box size of 8 fm ensures that the evolution of the charmonium wave function remains unaffected by boundary conditions.

Heavy quark potential is expected to be modified by thermal partons in the defined matter~\cite{Ali:2023kmr,Larsen:2024wgw,Luo:2024iwf,Liu:2024wez}. 
\tred{
The color screening effect reduces the real part of the heavy quark potential $ V_R  $ and the corresponding binding energies of heavy quarkonium. As a result, the mean radius of quarkonium eigenstates calculated using the screened potential increases with temperature~\cite{Satz:2005hx,Shi:2019tji,Song:2023zma}. The real part of the heavy quark potential is expected to lie between the free energy $ F(r,T) $ and the internal energy  $U(r,T)$~\cite{Du:2019tjf,Liu:2018syc,Liu:2010ej}. From the theoretical studies conducted by different research groups, it has been observed that the color-screened potential is closer to the limit of internal energy rather than free energy in heavy-ion collisions. When we take $V_R = U(r,T) $, the mean radius of charmonium increases slightly with $T$ when the temperature is close to $T_c $, and becomes divergent when $T$ is very large, indicating the "melting" of bound states. Consequently, the color screening effect on heavy quarkonium is minimal in the temperature regions just above \( T_c \), such as in the hot medium produced in p-Pb and pp collisions. 
Inspired by recent lattice QCD calculations~\cite{Bala:2021fkm} and phenomenological studies~\cite{Shi:2021qri}, we neglect the weak color screening effect in small collision systems and approximate the real part of the heavy quark potential with a Cornell potential~\cite{Wen:2022utn}, expressed as \( V_R = -\alpha/r + \sigma r \). The parameters are determined to be \( \alpha = \pi/12 \) and \( \sigma = 0.2 \ \text{(GeV)}^2 \), fitted to match the masses of charmonium eigenstates~\cite{Satz:2005hx,ParticleDataGroup:2018ovx}. Further studies are needed to assess the degree of color screening effects at high temperatures, particularly in the context of heavy-ion collisions.
}

\begin{figure}[!htb]
\includegraphics[width=0.45\textwidth]{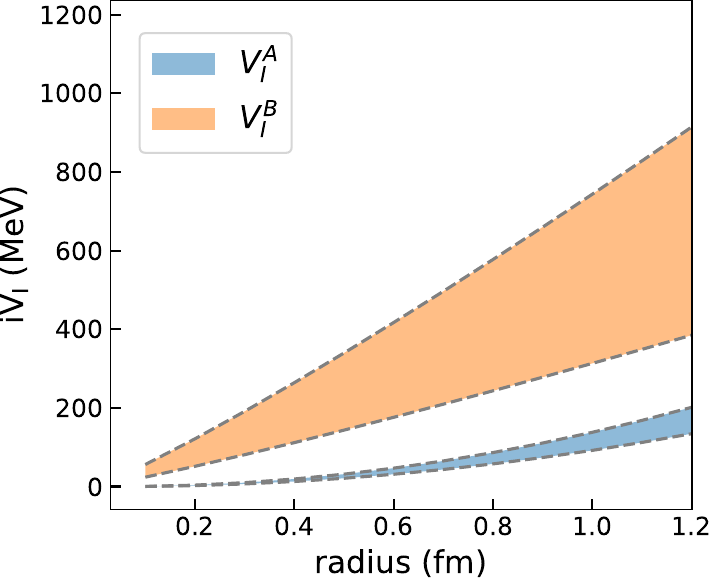}
\caption{Imaginary potential as a function of radius from two parameterizations. The blue band represents $V_I^{(A)}$, and the yellow band represents $V_I^{(B)}$. Each band's upper and lower edges correspond to temperature T=0.3 and 0.2 GeV, respectively.
}
\label{fig-lab-vi}
\end{figure}

 Besides, the parton inelastic scatterings contribute an imaginary part to the heavy quark potential. The imaginary part of the potential is connected with the real part~\cite{Larsen:2024wgw}. We directly fit the imaginary part of the potential given by lattice QCD calculations~\cite{Bala:2021fkm,Burnier:2016mxc}. Two kinds of parameterizations are performed to consider the uncertainty~\cite{Chen:2024iil,Wen:2022utn}, 
\begin{align}
    V_{I}^{(A)}(r,T)&=-iT(a_1 \bar r + a_2 \bar r^2),\\
    V_{I}^{(B)}(r,T)&=-iT[(rT)^{b_1}  + b_2 (rT)],
\end{align}
where $i$ is the imaginary number. In the first parametrization, the values of the parameters are fitted to be $a_1=-0.040$ and $a_2=0.50$. $\bar r\equiv r/fm$ is a dimensionless variable, $T$ is the temperature. In the second parametrization, $b_1=1.2$ and $b_2=0.54$. 
\tred{
In small collision systems, the typical temperatures of the medium are expected to be lower than those observed in nucleus-nucleus collisions. To illustrate the difference between the two imaginary potentials in the relevant temperature regions for pp collisions, we plot the two imaginary potentials at 0.2 GeV and 0.3 GeV. These values are slightly above \( T_c \) but below the typical temperatures encountered in Pb-Pb collisions.
}
Two parameterizations of the imaginary potential are plotted in Fig.\ref{fig-lab-vi}, ( T=0.2 GeV for the lower limit of each band) and (0.3 GeV for the upper limit of each band), respectively. These temperature values are expected to approximate the initial temperature ranges of the hot medium produced in high-multiplicity pp and p-Pb collisions. According to recent HotQCD calculations represented by $V_I^{(B)}$, the imaginary potential is higher than earlier estimates depicted by $V_I^{(A)}$. Both imaginary potentials will be incorporated into the Schr\"odinger equation to compute the suppression of charmonium. Despite the substantial uncertainties in the imaginary part of the potential, the ratio  
$\sigma_{\psi(2S)}/\sigma_{J/\psi}$ of $J/\psi$ is less affected. 
Our conclusion regarding the presence of small deconfined matter in high-multiplicity pp collisions remains unaffected.

Heavy quark pairs are produced in parton hard scatterings with a tightly bound wave function. When the size of the wave packet is small, the internal evolution of the wave packet is minimally affected by the medium on short-distance scales. As time progresses, the wave packet expands to a finite size where the effects of the hot medium become more pronounced. We assume that the hot medium reaches local equilibrium at a timescale of $\tau_0 = 0.4$ fm/c and employ the Schr\"odinger equation to evolve the charmonium wave packet from this initial time point. The initial wave function can be considered as a Gaussian function or charmonium eigenstates. In this study, we initialize the quarkonium wave packet at $\tau = \tau_0$ with different charmonium eigenstates $\psi(r,\tau_0) = r \Phi_{nl}(r)$.
Hot medium effects modify the wave package from this time point. When charmonium exits the QGP droplet, the final fractions $c_{nl}(t)$ of each charmonium eigenstate with quantum numbers $(n,l)$ can be determined via projection: $c_{nl} = \int \Phi_{nl}(r) \psi(r,t_f) r dr$. Here, $\psi(r,t_f)$ is governed by the Schr\"odinger equation Eq. (\ref{fun-rsch}), and $t_f$ denotes the time when the quarkonium moves to a region where the local temperature drops below the critical temperature  $T_c = 0.16$ GeV.

\begin{figure}[!htb]
\includegraphics[width=0.45\textwidth]{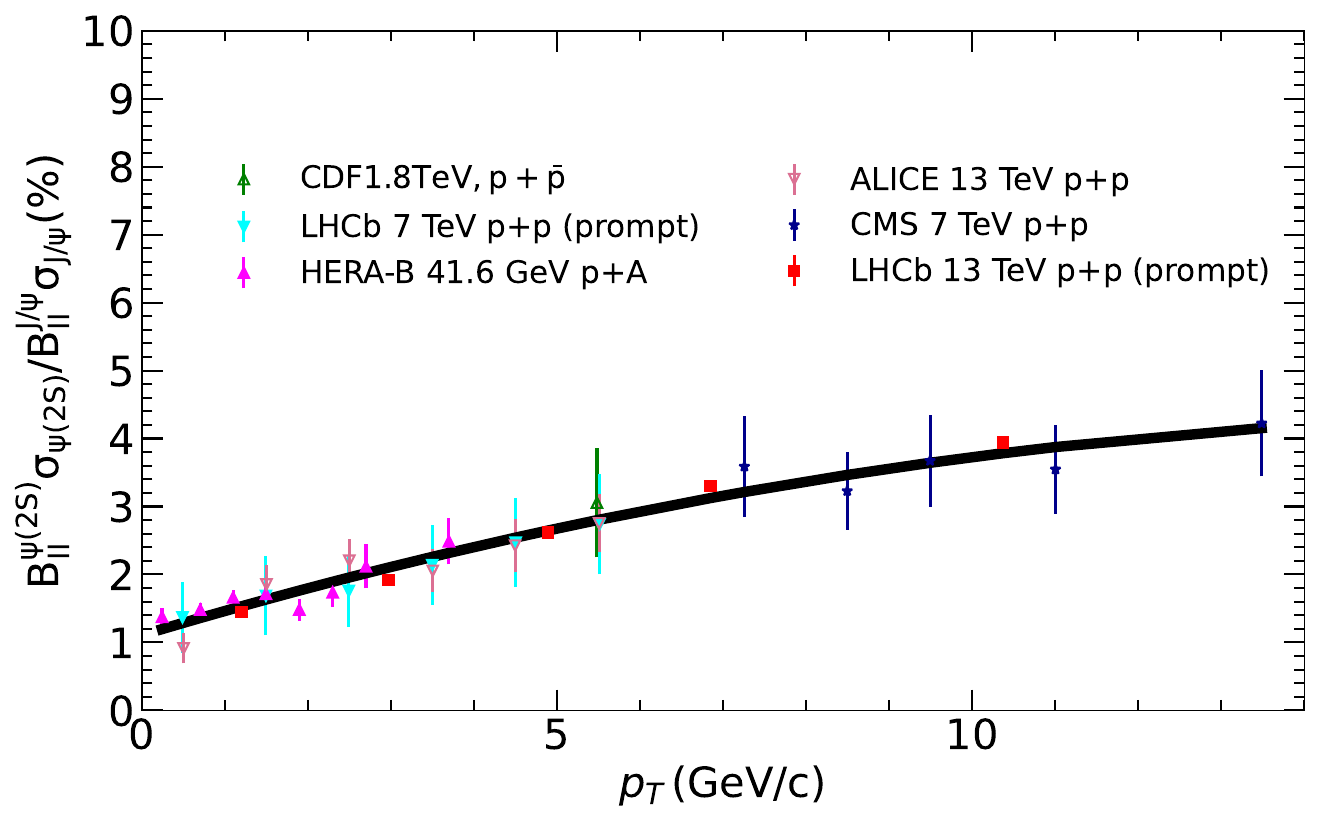}
\caption{Charmonium production ratio as a function of the transverse momentum $p_T$. The experimental data $\mathcal{B}_{ll}^{\psi(2S)}\sigma_{\psi(2S)}/(\mathcal{B}_{ll}^{J/\psi}\sigma_{J/\psi})$ are from ~\cite{LHCb:2023xie,PHENIX:2016vmz}. $\mathcal{B}_{ll}^{J/\psi}$ are the branching ratio of $J/\psi$ decaying into the dilepton, taken from Particle Data Group~\cite{ParticleDataGroup:2018ovx}. The line is the parametrization. 
}
\label{fig-ppT}
\end{figure}

To consider the finite size of the hot medium generated in pp collisions, one needs to consider the geometry size of a proton. A straightforward approximation is a uniform distribution of proton mass. In analogy to the nuclear distribution, we employ a similar form to describe the proton mass distribution~\cite{Chen:2016dke}, which avoids discontinuities at the proton edge, 
\begin{align}
    \rho_p = {\rho_0\over 1+ e^{(r-r_p)/a_p}}.
\end{align}
Inspired by the proton radius~\cite{Miller:2018ybm}, we estimate the value of parameters to be around $r_p=0.9$ fm and $a_p=0.1$ fm. $\rho_0=0.29\ \rm(fm)^{-3}$ is fitted with $\int d^3{\bf r}\rho_p=1$. In analogy to the case of nuclear collisions, the initial transverse distribution of the quarkonium is proportional to the function $T_p({\bf x}_T+{\bf b}/2)T_p({\bf x}_T-{\bf b}/2)$ where $T_p$ is the integration of the proton density along z-direction and ${\bf b}$ is the distance between the centers of two protons. The momentum distribution of the quarkonium satisfies a normalized distribution~\cite{Zhao:2020jqu,Chen:2018kfo}, 
\begin{align}
\label{eq-p-init}
{d^2N_\Psi \over d\phi p_T dp_T}={(n-1)\over \pi (n-2) \langle p^2_T\rangle} [1+{p_T^2\over (n-2) \langle p_T^2 \rangle}]^{-n}.
\end{align}
For the ground state $\Psi\equiv J/\psi$,
the mean transverse momentum distribution of $J/\psi$ without hot medium effects at 13 TeV is $\langle p_T^2 \rangle = 15$ $\rm (GeV/c)^2$. The value $n = 3.2$ is taken the same as the value at 5.02 TeV~\cite{Chen:2018kfo}. Initial momenta and positions of the quarkonium wave packet can be randomly generated based on the distribution given by Eq.(\ref{eq-p-init}) and the product of two proton thickness functions, $T_p({\bf x_T} - {\bf b}/2) T_p({\bf x_T} + {\bf b}/2)$. The local temperature of the medium used in the complex potential is given by the hydrodynamic model. By averaging many wave packages with different initial positions and transverse momenta, we determine the final fraction of the charmonium eigenstate ($n,l$) within the wave package. The survival probability of charmonium after moving out of the hot medium, is defined as the ratio of the final to initial fractions of the charmonium eigenstates, $S = \int d{\bf x_T}d{\bf p_T}|c_{nl}(t_f)|^2 / \int d{\bf x_T}d{\bf p_T}|c_{nl}(t_0)|^2$. For charmonium excited state $\psi(2S)$, the ratio of $\psi(2S)$ production cross section relative to the $J/\psi$ is given by experiments, shown as Fig.\ref{fig-ppT}.
The ratio varies with the transverse momentum. We parametrize the data points in Fig.\ref{fig-ppT} with the ansatz, 
\begin{align}
    f_{r}(p_T)=a_{r}p_T^2 + b_{r}p_T+c_{r}
\end{align}
with the value to be $a_{r}=-0.0103$, $b_{r}=0.365$, and $c_{r}=1.107$. This value is qualitatively consistent with previous values~\cite{PHENIX:2011gyb,ALICE:2021qlw}.
\tred{
Using the experimentally measured ratio \(\sigma_{\psi(2S)}/\sigma_{J/\psi}\), we can extract the cross sections for the direct production of \(J/\psi\) and \(\psi(2S)\) before the feed-down process. Finally, the survival probabilities \(R_{\psi}\) for \(J/\psi\) and \(\psi(2S)\) in the medium, corrected with the feed-down process \(\psi(2S) \rightarrow J/\psi\), can be obtained~\cite{Wen:2022utn,Wen:2022yjx,Islam:2020bnp}. In \(\sqrt{s_{NN}} = 13\) TeV pp collisions, experiments provide the normalized ratio \(\sigma_{\psi(2S),n}/\sigma_{J/\psi,n}\), defined as 
$
\frac{\sigma_{\psi(2S),n}/\sigma_{J/\psi,n}}{\sum_n \sigma_{\psi(2S),n}/\sum_n \sigma_{J/\psi,n}},
$
where \(n\) is the multiplicity bin index. We calculate the survival probabilities with the feed-down process, \(R_{J/\psi,n}\) and \(R_{\psi(2S),n}\), across different multiplicity bins labeled by \(n\). A similar normalization is performed:
$
\frac{R_{\psi(2S),n}/R_{J/\psi,n}}{\sum_n R_{\psi(2S),n}/\sum_n R_{J/\psi,n}},
$
which approximately equals to the normalized ratio \(\sigma_{\psi(2S),n}/\sigma_{J/\psi,n}\) given by experiments.
}

\begin{figure}[!htb]
\includegraphics[width=0.43\textwidth]{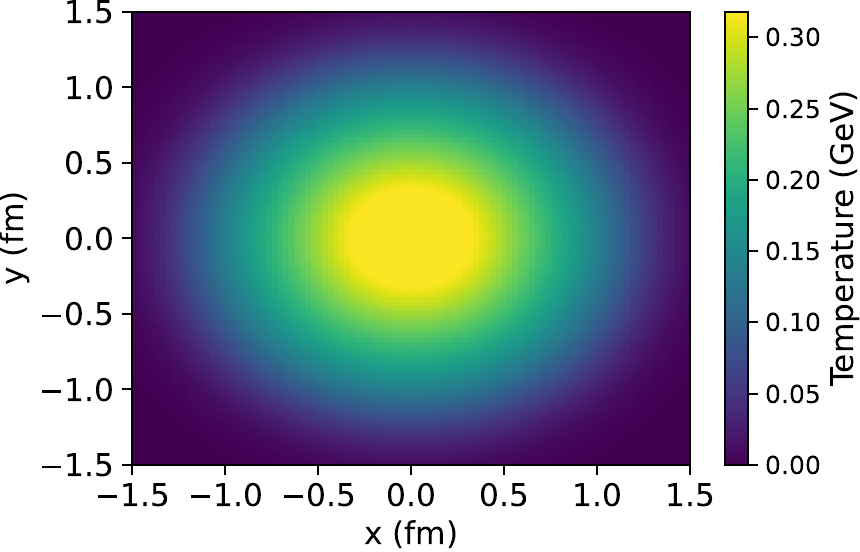}
\includegraphics[width=0.43\textwidth]{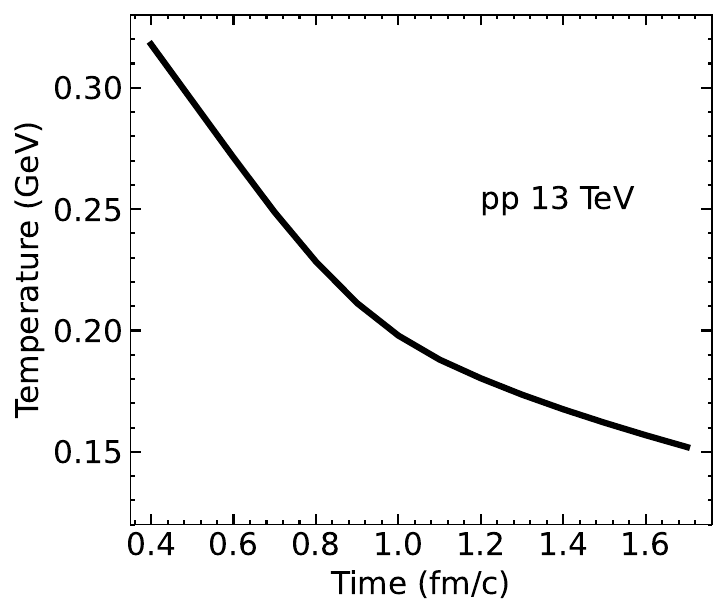}
\caption{(Upper panel) The spatial distribution of initial temperatures  $T(x,y)$ (GeV) of the QGP at $\tau_0 = 0.4$ fm/c in the most central pp collisions at $\sqrt{s_{NN}} = 13$ TeV.
(Lower panel) The time evolution of the temperature $T(t |{\bf x_T}=0)$ at the center of the QGP in pp collisions. 
}
\label{fig-temp-dis}
\end{figure}

In high-multiplicity pp collisions, to model the space-time evolution of the QGP droplets~\cite{PHENIX:2018lia}, we employ the widely-used hydrodynamic model (MUSIC) to generate temperature profiles~\cite{Schenke:2010rr,Schenke:2010nt}, which are subsequently utilized in the calculation of the Schr\"odinger equation. For the initial energy density, there may be significant fluctuations~\cite{Zhao:2022ugy}. We temporarily neglect the fluctuations of the medium and employ a smooth hydro for charmonium evolutions. The contribution of medium fluctuations on heavy quarkonium will be studied in detail in the following work. The initial energy density can be determined based on the final charged multiplicity observed in $\sqrt{s_{NN}}=13$ TeV pp collisions. This work attempts to quantify the hot medium effects with heavy quarkonium. We treat the initial temperature of the QGP as a parameter to be determined via charmonium. The maximal temperature of the smooth hydro is selected to be $T({\bf x_T}=0,\tau_0)=(0.2, 0.3, 0.4)$ GeV, respectively, in the most central pp collisions with $\tau_0=0.4$ fm/c. These different smooth hydro results will be used to study charmonium suppression and compared against experimental data. one temperature profile is plotted in Fig. \ref{fig-temp-dis}. The time evolution of temperatures at the center of QGP is plotted, with a lifetime of approximately 1-2 fm/c. This short lifetime and small volume of QGP in small collision systems result in weaker QGP suppression effects on charmonium~\cite{Chen:2018kfo,Wen:2022utn,Du:2018wsj} and bottomonium~\cite{Chen:2023toz,Wen:2022yjx,Du:2017qkv} compared to Pb-Pb collisions at similar colliding energies.

\section{Results in proton-proton collisions}

\begin{figure}[!htb]
\includegraphics[width=0.4\textwidth]{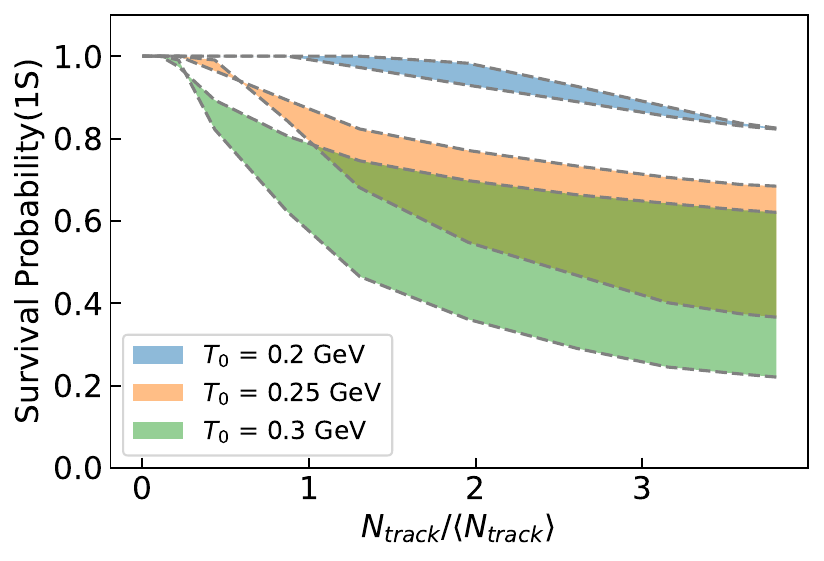}
\includegraphics[width=0.4\textwidth]{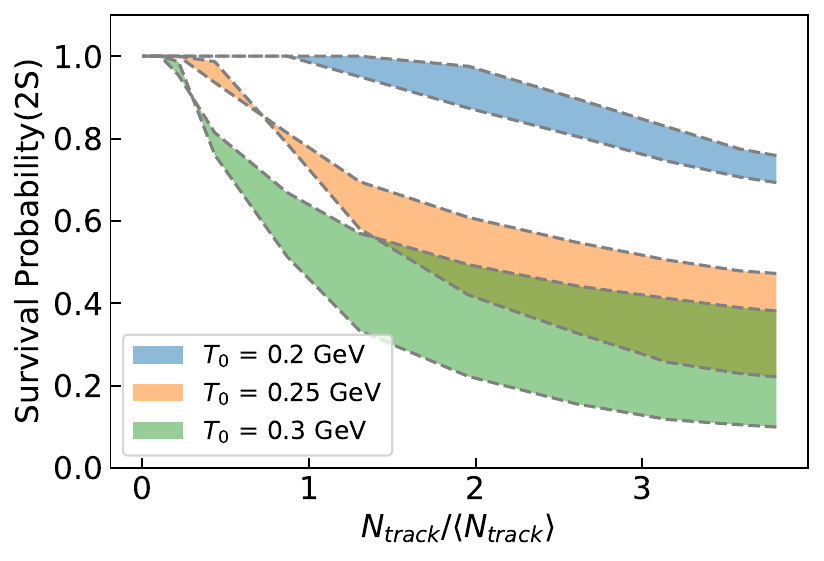}
\caption{The survival probability of $J/\psi$ (upper panel) and $\psi(2S)$ (lower panel) after moving out of the hot medium, as functions of the scaled number of tracks, which corresponds to the multiplicity bin in $\sqrt{s_{NN}}=13$ TeV pp collisions. The edges of the bands correspond to two imaginary potentials. Different bands correspond to different initializations of QGP with initial maximal temperatures $T_0({\bf x_T}=0)=0.2$, 0.25 and $0.3$ GeV, respectively.
}
\label{lab-fig-RAA}
\end{figure}

Utilizing the hydrodynamic model for QGP evolution and the time-dependent Schr\"odinger equation for charmonium evolution, one can calculate the nuclear modification factors of charmonium in pp collisions. Hot medium effects are encoded in the complex potential. Given the uncertainty in the imaginary part of the potential, we employ two parameterizations about previous lattice QCD calculations to calculate the direct nuclear modification factors of $J/\psi$ and $\psi(2S)$, plotted in Fig.\ref{lab-fig-RAA}. The upper and lower limits of the band correspond to the calculations with $V_{I}^{(A)}$ and $V_I^{(B)}$, respectively. 
Different hydrodynamic scenarios are considered, varying in initial temperatures. The upper panel of Fig. \ref{lab-fig-RAA} displays $J/\psi$ nuclear modification factors under different initialization of QGP. The x-axis represents the ratio of the number of tracks to its mean value, $N_{track}/\langle N_{track}\rangle$, which correlates with the particle multiplicity. In experiments, the number of tracks in pp collisions depends on pseudorapidity regions, with distinct values in forward and backward pseudorapidities denoted as $N_{fwd}$ and $N_{bwd}$, respectively. 
We employ three initializations of the QGP temperatures to calculate charmonium suppression and compare them with the experimental data in forward and backward pseudorapidities.

In Fig.\ref{lab-fig-RAA}, the different colors of the bands correspond to the calculations with different hydro. As there is a significant difference in the two parametrizations of imaginary potential, especially at high-temperature regions, the width of the band is also large in the case of a hotter medium labeled with $T_0(\tau_0)=0.3$ GeV. At the temperature T=0.3 GeV, the magnitude of the $V_{I}^{(B)}$ is around ten times larger than the value of $V_I^{(A)}$ at the distance $0.5$ fm of $J/\psi$ mean radius. This gives much stronger suppression on $J/\psi$ and $\psi(2S)$ yields than the case with $V_{I}^{(A)}$. The width of the band is very large. While at a lower temperature like T=0.2 GeV, the difference between $V_I^{(A)}$ and $V_I^{(B)}$ becomes smaller, where the width of the $R_{AA}$ band is also smaller (see the band with $T_0=0.2$ GeV in the figure). In the low multiplicity region, the hot medium effect is very weak. The nuclear modification factors of $J/\psi$ and $\psi(2S)$ approach unit.

\begin{figure}[!htb]
\includegraphics[width=0.45\textwidth]{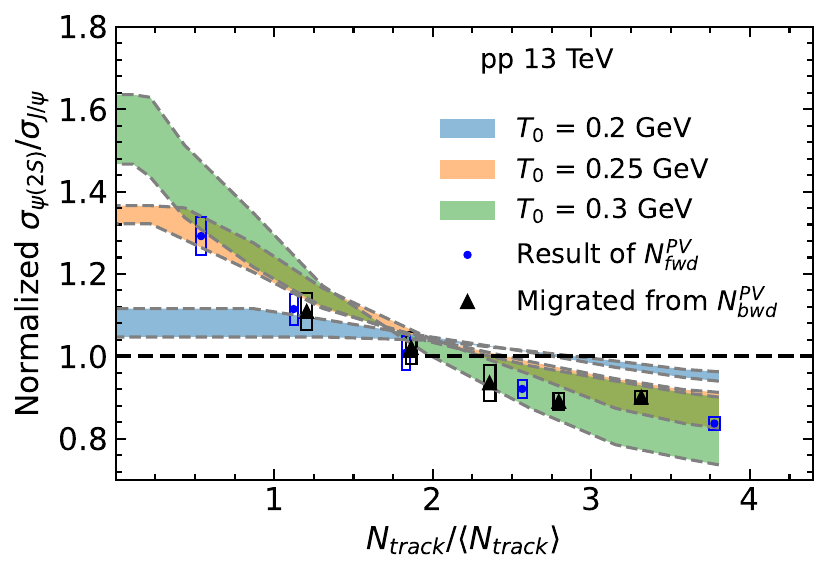}
\caption{Normalized ratio of charmonium production cross section $\sigma_{\psi(2S)}/\sigma_{J/\psi}$ as a function of $N_{track}/\langle N_{track}\rangle$ in $\sqrt{s_{NN}}=13$ TeV pp collisions. Three bands correspond to three kinds of hydro with the initial maximal temperature of $T_0=(0.3,0.25,0.2)$ GeV, respectively. The edges of each band correspond to two parameterizations of the imaginary part of the potential. The experimental data are selected in the forward and backward pseudorapidities from LHCb Collaboration~\cite{LHCb:2023xie}.
}
\label{lab-exp-ratio}
\end{figure}

Despite significant uncertainties in the nuclear modification factors of charmonium due to varying imaginary potentials, as depicted in Figs. \ref{fig-lab-vi} and \ref{lab-fig-RAA}, the ratio of $J/\psi$ and $\psi(2S)$ production cross sections exhibits lower uncertainty. Experimental data provide the normalized ratio $\sigma_{\psi(2S)}/\sigma_{J/\psi}$ as a function of the scaled number of tracks $N_{track}/\langle N_{trach}\rangle_{NB}$. Similarly, theoretical results are plotted with bands in Fig. \ref{lab-exp-ratio}. In the absence of hot medium effects, the normalized $\sigma_{\psi(2S)}/\sigma_{J/\psi}$ approaches unity and shows little dependence on multiplicity. However, at higher values of $N_{track}/\langle N_{track}\rangle$, corresponding to higher multiplicities, the effects of the hot medium on charmonium become more pronounced. Specifically, $\psi(2S)$ experiences greater suppression than $J/\psi$, particularly at larger distances where the magnitude of the imaginary potential increases. The upper and lower limits of the bands in Fig. \ref{lab-exp-ratio} represent scenarios using two distinct imaginary potentials, $V_I^{(A)}$ and $V_I^{(B)}$. To quantify the influence of medium temperatures, we incorporate different hydrodynamic profiles into the Schr\"odinger equation model, setting the maximum temperatures $T_0=0.2$, 0.25, and $0.3$ GeV for the most central pp collisions at $\sqrt{s_{NN}}=13$ TeV. The band with $T_0= 0.25$ GeV, effectively describes the normalized $\sigma_{\psi(2S)}/\sigma_{J/\psi}$. In contrast, lower medium temperatures lead to weaker suppression of charmonium, resulting in the ratio $\sigma_{\psi(2S)}/\sigma_{J/\psi}$ displaying a flatter trend with $N_{track}/\langle N_{track}\rangle$, as illustrated by the band with $T_0=0.2$ GeV in Fig.\ref{lab-exp-ratio}. 
The ratio $\sigma_{\psi(2S)}/\sigma_{J/\psi}$ in Fig.\ref{lab-exp-ratio} is already normalized. It is almost unaffected by the input of the $J/\psi$ and $\psi(2S)$ production cross sections and the cold nuclear matter effects. 
The distinct separation between these bands effectively demonstrates the impact of the hot medium on charmonium. Notably, the uncertainties associated with the complex heavy quark potential do not influence these conclusions.

\begin{figure}[!htb]
\includegraphics[width=0.4\textwidth]{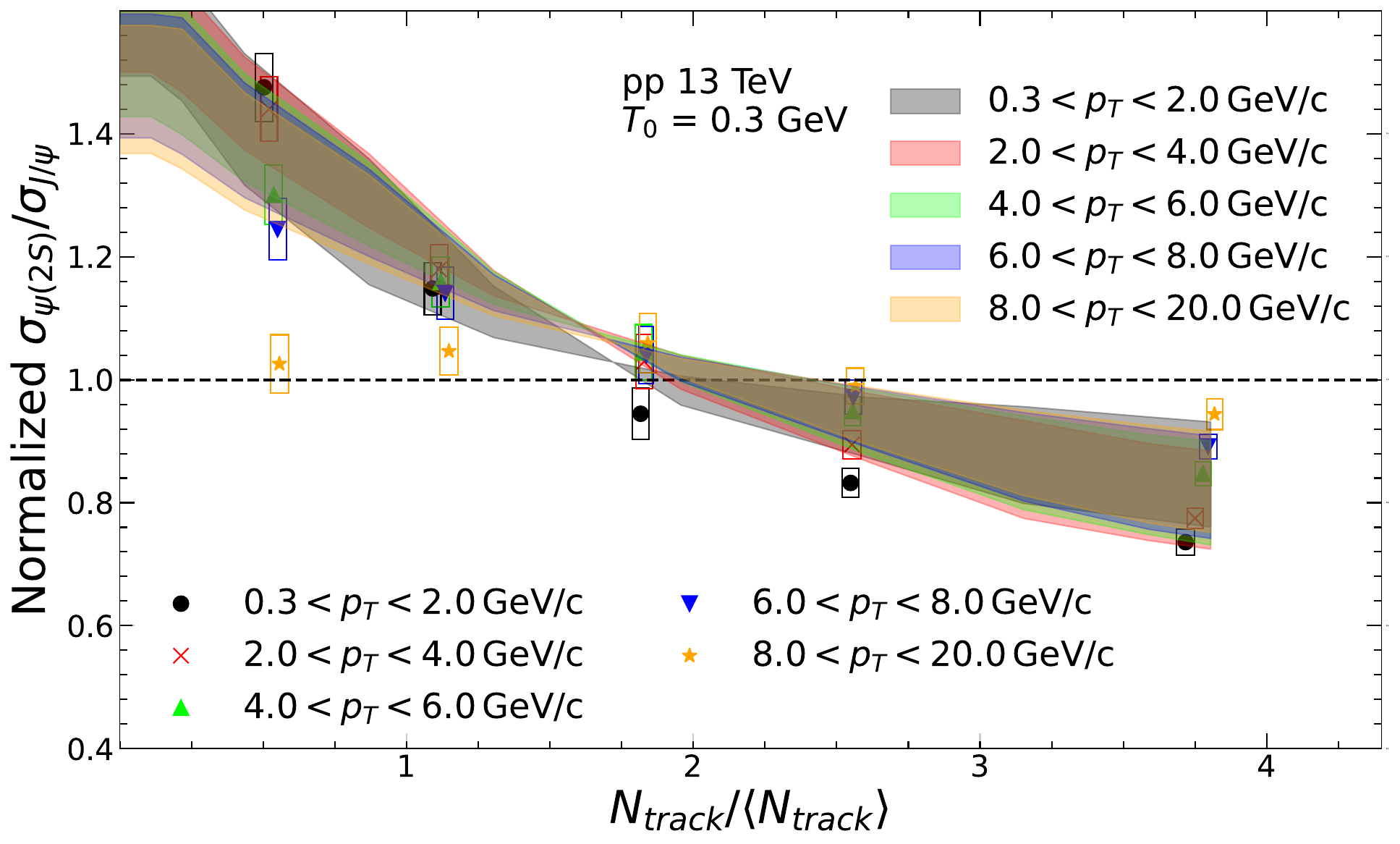}
\includegraphics[width=0.4\textwidth]{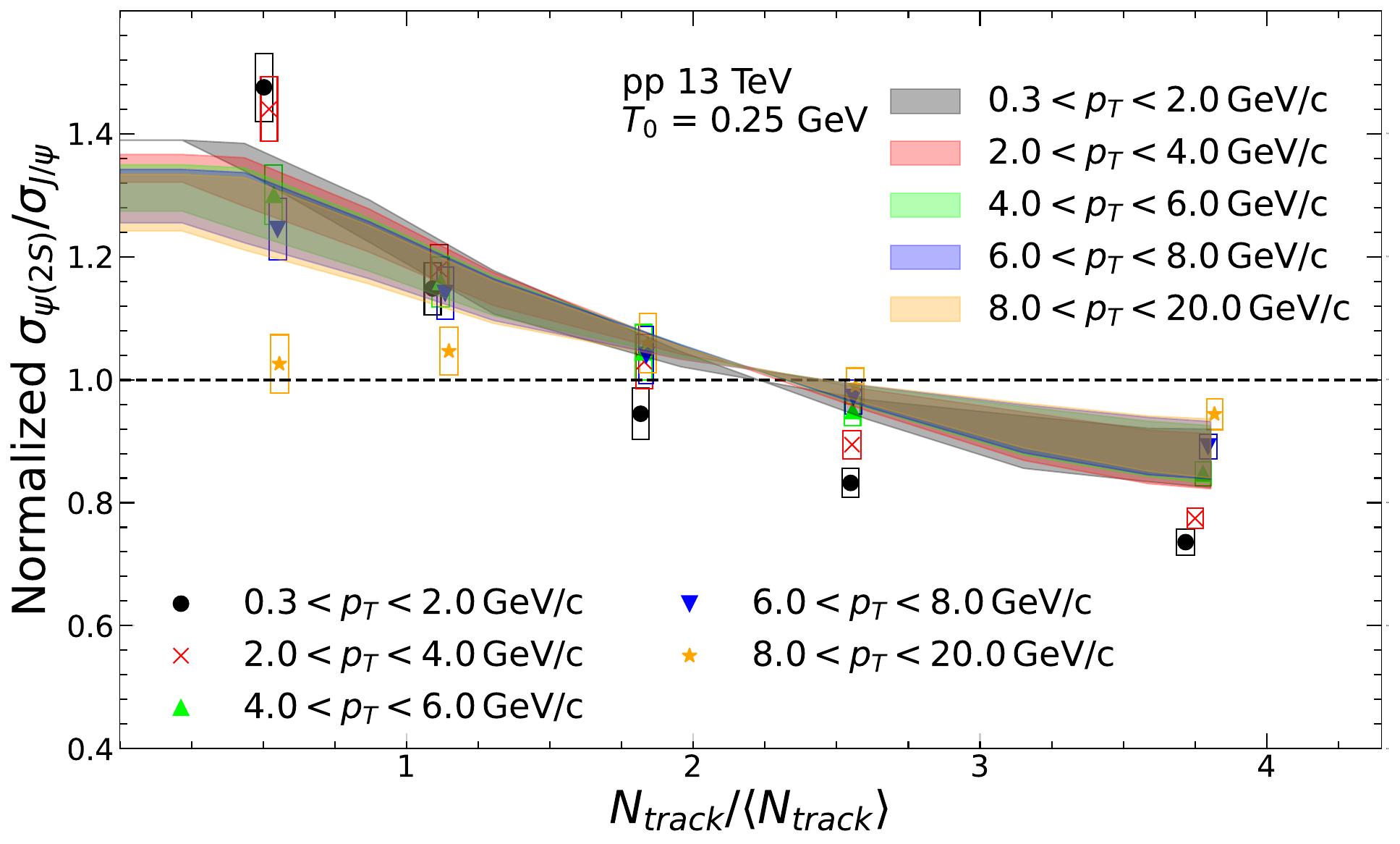}
\includegraphics[width=0.4\textwidth]{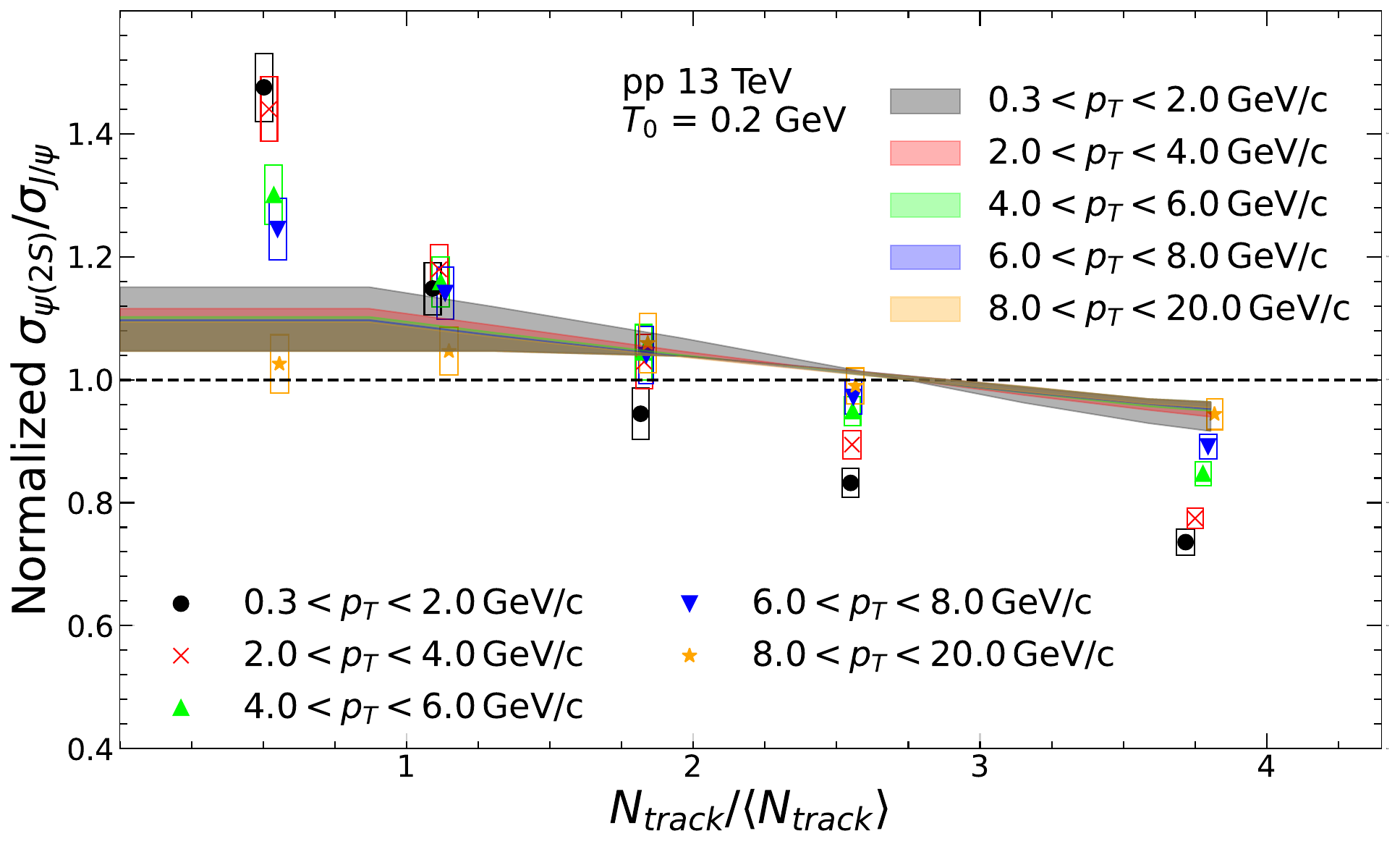}
\caption{The normalized ratio of charmonium production cross section $\sigma_{\psi(2S)}/\sigma_{J/\psi}$ in different $p_T$ bins as a function of $N_{track}/\langle N_{track}\rangle$ are plotted in $\sqrt{s_{NN}}=13$ TeV pp collisions. Experimental data is from the LHCb Collaboration~\cite{LHCb:2023xie}. Different temperature profiles are employed to calculate charmonium suppression, with initial maximal temperatures set as $T_0({\bf x_T}=0)=0.3$ GeV (upper panel), 0.25 GeV (middle panel), and 0.2 GeV (lower panel), respectively.
}
\label{lab-ptbins}
\end{figure}

The ratio of charmonium production cross-sections is also measured in different $p_T$ bins. In Fig. \ref{lab-ptbins}, we calculate the normalized $\sigma_{\psi(2S)}/\sigma_{J/\psi}$ using different initializations of QGP droplets, where the maximum initial temperature of QGP is set to $T_0({\bf x_T}=0)=0.3$ GeV (upper panel), 0.25 GeV (middle panel), and 0.2 GeV (lower panel), respectively.
With increasing transverse momentum, the quarkonium exits QGP droplets more quickly, resulting in slightly weaker QGP suppression. The value of $\sigma_{\psi(2S)}/\sigma_{J/\psi}$ approaches unity in higher $p_T$ bins. The tendency of theoretical calculations across different $p_T$ bins is consistent with the experimental data, with less noticeable variation than the experimental results.
Calculations using a medium initial temperature of $T_0({\bf x_T}=0)=0.2$ GeV underestimate the experimental data, whereas theoretical results with higher initial temperatures of $T_0({\bf x_T}=0)=0.25$ GeV and 0.3 GeV quantitatively match the data. It should be noted that employing a precise imaginary potential with reduced uncertainty would narrow the width of the bands in the figure.

\section{Summary}
We utilize the Schr\"odinger equation and the hydrodynamic model to investigate the evolution of charmonium and QGP droplets in high multiplicity proton-proton collisions at $\sqrt{s_{NN}}=13$ TeV. Hot medium effects primarily alter the charmonium yield through parton inelastic scatterings encoded in the complex heavy quark potential. In contrast, the color screening effect on heavy quarkonium is expected to be minimal.
Due to uncertainties in the imaginary part of the heavy quark potential, two different parameterizations of the potential are employed to compute the nuclear modification factors of charmonium and the ratio of normalized charmonium production cross-sections, $\sigma_{\psi(2S)}/\sigma_{J/\psi}$. The value of the normalized $\sigma_{\psi(2S)}/\sigma_{J/\psi}$ is independent of initial effects before the production of heavy quark pair and is primarily affected by hot medium effects.
The significant suppression of the normalized $\sigma_{\psi(2S)}/\sigma_{J/\psi}$ in high multiplicity bins of proton-proton collisions is interpreted as a clear signal of final-state interactions, where the larger geometrical size of $\psi(2S)$ compared to the ground state $J/\psi$ leads to more significant suppression by the hot medium. The Schr\"odinger model with a complex potential well describes the experimental data of normalized $\sigma_{\psi(2S)}/\sigma_{J/\psi}$. The ratio in different $p_T$ bins is also calculated and quantitatively explains the trend observed in experimental data, where charmonium with higher $p_T$ experiences weaker suppression due to their shorter escape time from QGP droplets.

\vspace{1cm}
{\bf Acknowledgement:} We appreciate Jiamin Liu's assistance and discussions regarding the hydrodynamic model.
This work is supported by the National Natural Science Foundation of China
(NSFC) under Grant Nos. 12175165. 




\end{document}